\documentclass[aps, prl,
                   twocolumn,
]{revtex4}

\usepackage{graphicx}
\usepackage{color}
\usepackage{epsfig}
\usepackage{amsmath}
\usepackage{amstext}
\usepackage{amssymb}
\usepackage{bm}

\usepackage[latin2]{inputenc}

\DeclareGraphicsExtensions{.eps,.ps}
\newcommand{\CNOT}{\rm CNOT}
\newcommand{\SWAP}{\rm SWAP}
\newcommand{\SWAT}{\rm SWAT}
\newcommand{\I}{\rm I}

\begin{document}

\author{Antoni W{\'o}jcik }
\email{antwoj@amu.edu.pl}

\author{ Ravindra W. Chhajlany }
\email{ravi@amu.edu.pl}

\affiliation{Faculty of Physics, Adam Mickiewicz University,
 \\ Umultowska 85,  61-614 Pozna{\'n}, Poland}
 \date{\today}

\date{\today}
\title{Quantum-classical correspondence in the oracle model of computation}

\begin{abstract}
The oracle model of computation is believed to allow a rigorous proof
of quantum over classical computational superiority.  Since quantum
and classical oracles are essentially different, a correspondence
principle is commonly implicitly used as a platform for comparison of
oracle complexity. Here, we question the grounds on which this
correspondence is based. Obviously, results on quantum speed-up depend
on the chosen correspondence. So, we introduce the notion of genuine
quantum speed-up which can serve as a tool for reliable comparison of
quantum vs classical complexity, independently of the chosen
correspondence principle.
\end{abstract}


\maketitle

Quantum mechanics offers a new, promising perspective for computer
science. Quantum computers are believed to hold a computational
advantage over classical ones. One of the most spectacular examples of
this quantum speed-up is Shor's famous algorithm \cite{shor} for
factorisation of large numbers. The executing time of Shor's algorithm
scales polynomially with the size of the problem, whereas the best
known classical algorithm - General Number Field Sieve scales
subexponentially. Of course the notion of quantum speed-up is not
absolute unless it is judged by comparing optimal quantum and
classical algorithms. However, finding lower bounds for NP problems is
not easy in general. Thus, in order to prove the advantage of quantum
computers in a rigorous way a special model of computation, namely the
oracle model of computation (OMC) was introduced. In the OMC,
algorithmic complexity is identified with query complexity, {\it i.e.}
the number of oracle calls required for solving a problem. Within this
model, quantum and classical bounds for many problems have been
obtained.  The cases when quantum complexity is lower than
classical complexity are usually claimed to be rigorous proofs of
quantum speed-up. Obviously, algorithms which are to be compared
within the OMC should call the same oracle. Quantum and classical
oracles are essentially different.  Thus, strictly speaking, reliable
comparison of quantum and classical algorithms is not possible. To
overcome this problem the notion of correspondence between quantum and
classical oracles is commonly used.

% It is a commonly accepted method to compare a quantum oracle with its
% corresponding classical oracle.

In this Letter, we show that this correspondence can not in general be
unique. As a consequence, we propose a modified procedure for reliable
comparison of quantum and classical algorithms within the OMC. Within
this framework it turns out that quantum speed-up offered by some
algorithms is just an artefact of the ambiguity of the previously used
correspondence.  Our arguments also shed some light on the role of
entanglement in quantum speed-up.

Let us consider the question of ``quantizing'' a given classical
operation. As an example to clear notions, suppose the classical
operation is the (one bit) NOT gate which converts a bit ($a$) into
its compliment ($1-a$), ($a=0,1$). It seems natural to choose as a
quantum counterpart of this gate the $\sigma_{x}$ Pauli operator
\begin{gather}
\sigma_{x}= \left(
\begin{array}{lr}
0   & 1   \\
1   & 0
\end{array}
\right).
\label{Ra3a}
\end{gather}
Indeed, the transformation invoked by this matrix on choosing the quantum
 bit in the form $\chi_{a}= |a \rangle \langle a | $, where $|a
 \rangle = \left(
\begin{array}{c}
1-a       \\
  a 
\end{array}
\right)$ is exactly 
\begin{gather}
\chi_{a} \xrightarrow{\sigma_{x}} \chi_{1-a}.
\label{Rb3a}
\end{gather}
On the other hand, the $\sigma_{z}$ Pauli operation 
\begin{gather}
\sigma_{z} = \left(
\begin{array}{lr}
1   & 0   \\
0   &-1
\end{array}
\right)
\label{Rc3a}
\end{gather}
also implements the NOT gate 
\begin{gather}
\eta_{a} \xrightarrow{\sigma_{z}}
\eta_{1-a}
\label{Rs3a}
\end{gather}
 provided that the computational basis states are chosen
in a different way, namely $\eta_{a}= |a \rangle '' \langle a | $,
where $|a \rangle' = \frac{1}{\sqrt{2}}\left(
\begin{array}{c}
     1  \\
   (-1)^{a}
\end{array}
\right) $. This simple example illustrates the dependence of the
quantum-classical correspondence on the choice of computational basis
states. Furthermore, even after choosing a given computational basis
there remains  a different ambiguity described below. Suppose we
choose the bit $\chi_{a}$, then the operation $\tilde{\sigma }_{x}$
\begin{gather}
\tilde{\sigma }_{x} = \left(
\begin{array}{lr}
0   & e^{i \theta }\\
e^{i \phi }& 0
\end{array}
\right)
\label{Re3a}
\end{gather}
implements the bit compliment
\begin{gather}
\chi_{a}
\xrightarrow{\tilde{\sigma}_{x}} \chi_{1-a}
\label{Ru3a}
\end{gather}
 for any $\theta ,\phi $. Let us emphasize that the common convention
of choosing $\theta = \phi =0$ is arbitrary and can not be justified
simply by correspondence. This is because  physical
states  are represented by rays not vectors in Hilbert space.

The argument given above is bidirectional, {\it i.e.} to a given  quantum
operation there may correspond different classical operations. For
example, the $\sigma_{z}$
operation corresponds to the classical identity 
\begin{gather}
\chi_{a} \xrightarrow {\sigma_{z}} \chi_{a}
\label{Rg3a}
\end{gather}
 as well as NOT operation
\begin{gather}
\eta_{a} \xrightarrow{\sigma_{z}} \eta_{1-a}.
\label{Rh3a}
\end{gather}

In general, we say that a classical reversible operation $O$
transforming m-bits into m-bits according to the rule 
\begin{gather}
  \vec{a}
  \xrightarrow{O} \vec{b} ,
\label{Ry3a}
\end{gather}
where $\vec{z}= (z_{1},z_{2} , \ldots, z_{m}) , \vec{z}=\vec{a},\vec{b}$, corresponds to
  a quantum operation $U$ transforming m-qubits into m-qubits iff
\begin{gather}
\rho_{\vec{a}} \xrightarrow{U} \rho_{\vec{b}}
\label{Rz3a}
\end{gather}
  for  $\rho _{\vec{z}} = \rho_{z_{1}}^{(1)}\otimes \rho_{
  z_{2}}^{(2)}\otimes \ldots \otimes \rho_{z_{m}}^{(m)}$.  This
  correspondence is based on the formal identification of the action
  of the classical and quantum operations on their respective
  computational states. Note that each of the single qubit quantum
  computational states may be chosen arbitrarily and may be different
  for each qubit as long as they are pure and satisfy the
  orthogonality condition
  $\rho_{z_{j}}^{(j)}. \rho_{1-z_{j}}^{(j)}=0$.

Let us now consider the so-called standard oracle which is a $n+1$-qubit
unitary operation defined below
\begin{gather}
  |\vec {x} \rangle |y \rangle \xrightarrow{U_{S}^{f}} |\vec{x} \rangle
  |y \oplus f(\vec{x}) \rangle ,
  \label{Ri3a}
\end{gather}
where $|\vec{x}\rangle = \otimes_{j} |x_{j}\rangle^{(j)}$ denotes an
$n$-bit quantum register ($x_{j},y = 0,1$) and $f:\{0,1\}^{n} \mapsto
\{0,1 \}$. On choosing the computational states of the form
$\rho_{x_{j}}^{(j)} = \chi_{x_{j}}, \rho_{y}= \chi_{y}$, the action of
the oracle is
\begin{gather}
  \rho_{\vec{x}} \otimes \rho_{y} \xrightarrow{U_{S}^{f}} \rho_{\vec{x}}
  \otimes \rho_{y \oplus f(\vec{x})},
  \label{Rj3a}
\end{gather}
where $\rho_{\vec{x}}= \otimes_{j=1}^{n} \rho_{x_{j}}$. Thus, a natural 
corresponding classical counterpart  $O_{S}^{f}$ of this oracle transforms $n+1$
bits as follows:
\begin{gather}
(\vec{x}),(y) \xrightarrow{O_{S}^{f}} (\vec{x}),(y \oplus
  f(\vec{x})), 
\label{Rk3a}
\end{gather}
where $\vec{x}= (x_{1}, x_{2}, \ldots ,x_{n})$.

As explained earlier, this classical oracle need not be a unique
counterpart of $U_{S}^{f}$. Indeed, by choosing a different set of
computational states of the form $\rho_{x_{1}}^{(1)}= \eta_{x_{1}}$,
$\rho_{x_{j}}^{(j)}= \chi_{x_{j}} $ ($j \neq 1$), $ \rho_{y}=
\eta_{y}$ the quantum oracle implements the transformation
\begin{gather}
  \rho_{\vec{x}} \otimes \rho_{y} \xrightarrow{U_{S}^{f}}
  \rho_{\vec{x} \oplus \vec{c}}
  \otimes \rho_{y},
\label{Rm3a}
\end{gather}
where $\vec{c}= (c,0, \ldots, 0)$ and $ c= f(0, x_{2}, \ldots, x_{n}) \oplus f(1, x_{2}, \ldots
,x_{n})$. The classical counterpart corresponding to this
transformation of computational states is now $O_{A}^{f}$
\begin{gather}
(\vec{x}), (y) \xrightarrow{O_{A}^{f}}
  (\vec{x} \oplus \vec{c}),  (y)  .
\label{Rn3a}
\end{gather}

So, there are at least two mainfestly different classical oracles
corresponding to the standard quantum oracle. It is interesting to see
what effect this ambiguity in the quantum classical correspondence has
on the quantum speed-up of oracle problems. Let us focus on the well
known PARITY problem \cite{parity} which generalizes the original Deutsch
problem \cite{deutsch}. In the oracle setting, this problem requires deciding whether
$ \sum_{\vec{x}}^{} f(\vec{x})$ is even or odd. The optimal
classical algorithm to the standard classical oracle $O_{S}^{f}$
requires $N=2^{n}$ queries to solve this problem, whereas it suffices
to query the quantum oracle $U_{S}^{f}$ just $N/2$ times. Hence, the
quantum speed-up exhibited by $U_{S}^{f}$ as compared to $O_{S}^{f}$
is simply by a constant factor. On the other hand, the classical oracle
$O_{A}^{f}$ requires exactly the same number of queries $N/2$ as the
quantum oracle. Thus, there is no quantum speed-up at all when the
oracles $U_{S}^{f}$ and $O_{A}^{f}$ are compared.

As another example, consider the slightly modified Bernstein-Vazirani
 problem.  The oracle function $f:\{0,1\}^{n} \mapsto \{0,1\}$ is
 promised to be of the form $f(\vec{x}) = k_{0} + \vec{k}\cdot
 \vec{x}$, where $\vec{k} = (k_{1}, k_{2} , \ldots , k_{n})$ is the
 $n$-bit string to be identified. Notice that both the Deutsch problem
 and the original Bernstein-Vazirani (BV) problem \cite{bv} are
 special cases of the stated problem when $n=1$ and $k_{0}=0$
 respectively. Comparison of the standard classical oracle $O_{S}^{f}$
 which optimally requires $n+1$ calls with the quantum oracle
 $U_{S}^{f}$ which requires only a single call yields linear quantum
 speed-up.  Now, let us choose a different set of computational states
 for the quantum oracle of the form $\rho_{x_{j}}^{(j)} =
 \eta_{x_{j}}$ (for all $j$) and $\rho _{y}= \eta_{y}$. The action of
 the quantum oracle on these states is
\begin{gather}
  \rho_{\vec{x}} \otimes \rho_{y} \xrightarrow{U_{S}^{f}} \rho_{\vec{x}+\vec{k}}
  \otimes \rho_{y}.
\label{Rv3a}
\end{gather}
Hence, the classical oracle corresponding to this transformation is $O_{B}^{f}$
\begin{gather}
(\vec{x}), (y) \xrightarrow{O_{B}^{f}}
  (\vec{x} \oplus \vec{k}),  (y) .
\label{Rw3a}
\end{gather}
Obviously a single call to the classical oracle $O_{B}^{f}$ suffices
to solve the promise problem. Again, we conclude that there exists a
classical oracle corresonding to the quantum oracle which is just as
efficient.

Let us now comment on the interpretation of the above simple
examples. In the usual scenario, one starts with the standard
classical oracle which is then replaced by its quantum
counterpart. Our results do not question the fact that the quantum
oracle may provide a more efficient solution to a formulated oracular
problem than the standard classical oracle.  However, it is the
algorithms not that the oracles that should be compared.  As mentioned
in the introduction, since quantum and classical oracles are
completely different, strict comparison of the algorithms that call
these oracles is meaningless. If indeed this comparison is made, then
the source of the advantage of the better ``quantum'' solution could
be hidden within the quantum oracle itself, although it may seem to be
manifested in the quantum algorithm. Indeed providing the quantum
oracle may be equivalent to providing different (non-standard)
classical oracles.

To clarify further the notion of quantum speed-up in the OMC, suppose
Alice and Bob (who is constrained to use only classical operations on
logical bits) compete with each other to get a quicker solution to a
given oracular problem.  Suppose both Alice and Bob are given the same
classical device (oracle). In this case, Alice cannot use quantum
mechanical operations to her advantage since one cannot construct a
quantum oracle given a closed classical black-box. Now suppose both
Alice and Bob are given the same quantum oracle. Quantum speed-up
occurs when Alice can provide a more efficient solution than Bob (this
is indeed the case, {\it e.g.} in Grover's search algorithm).  In the
BV problem however, Alice will manage a quicker solution only if Bob
is additionally forced to use a particular (inefficient) encoding of
logical states. Suppose the quantum oracle is implemented by an
optical system closed in a black box whose input and output ports
consist of optical fibres. Assume the logical bits to be encoded in
the polarization of light.  The classical nature of Bob's state
implies that he can use only two orthogonal polarization states, {\it
e.g.} vertical and horizontal. Notice that the number of steps Bob
needs to solve the problem ($n$ or $1$) depends just on the
orientation of the device ($0^{o}$ or $45^{o}$ respectively).

% This is usually
% assumed to exemplify the advantage of the quantum oracle.

% It must be emphasized that the competition between classical and
% quantum computation should be based on algorithms n
% query complexity should  measure the algorithmic
% complexity relative to equivalent oracles.

Thus, we pose the question whether the speed-up in oracle 
problems is genuine quantum speed-up or just the result of the
interplay between two classical oracles.  Answering this question
requires a refined procedure of comparing quantum and classical
oracles.  Here, we postulate the detection of genuine quantum speed-up
by comparing the quantum oracle to its best possible corresponding
classical counterpart.

Our considerations also resolve the apparent puzzle of ``infinite''
quantum speed-up in BV algorithms. From Eqs.(\ref{Rv3a}) and
(\ref{Rw3a}), notice that the query bit is not transformed at all.
Therefore, especially in experimental realizations \cite{exp}, the
query bit is completely excluded and the BV circuit is implemented as
a controlled-$f$ phase shift oracle,
\begin{gather}
|\vec{x} \rangle \xrightarrow{U_{f}} (-1)^{\vec{x} \cdot \vec{k}} |\vec{x} \rangle 
\label{Rb4a}
\end{gather}
The standard classical counterpart of this oracle does not allow the
 extraction of any information about $\vec{k}$, since it is simply the
 Identity oracle:
\begin{gather}
(\vec{x}) \xrightarrow{O^{f}_{S}} (\vec{x}).
\label{Rc4a}
\end{gather}
 Since the quantum oracle recovers the value of $\vec {k}$ in a single
 query, it would seem that there is ``infinite'' quantum speed-up for
 this oracle setting.  However, notice that there exists a different
 classical counterpart 
\begin{gather}
(\vec{x}) \xrightarrow{O^{f}_{\tilde{B}}} (\vec{x} \oplus \vec{k})
\label{Rd4a}
\end{gather}
which also recovers $\vec{k}$ in a single call and thus resolves the 
puzzle.

Below, we sketch the general problem of finding all possible classical
counterparts of an arbitrary unitary and solve it for the simplest
case of 2-qubit unitaries.  A general reversible classical oracle
acting on $m$-bits is a permutation $O$ of all $2^{m}$ possible input
strings.  On the other hand, a quantum oracle is of course a general
$m$-qubit unitary operation $U$. When does a classical oracle $O$
correspond to a quantum oracle $U$? As discussed in the introductory
part of this Letter (see Eq.(\ref{Re3a})), for characterizing classical counterparts, one
must consider generalized permutation unitaries $P$ whose non-zero
entries are unit modulus complex numbers. We say that a unitary matrix
$U$ has a classical counterpart $O$ (in accordance with
Eqs.(\ref{Ry3a}) and (\ref{Rz3a})) iff $U$ is locally equivalent to
$P= D O$, {\it i.e.}
\begin{gather}
(\otimes_{i} L^{(1)}_{i}) U (\otimes_{i} L^{(2)}_{i}) = P,
\label{Rx3a}
\end{gather}
where all $L^{(1)}_{i}, L^{(2)}_{i}$ are single qubit operations and $D$ is a
diagonal unitary matrix. The problem of finding all possible classical
counterparts is a particular subset of the general problem of local
equivalence of unitary operations.  Unfortunately, no general solution
to this problem has been obtained so far.

In the simplest case of $2$-qubit unitaries $U$, three real parameters
completely characterize local equivalence \cite{makhlin, cirac,
zhang}. A computationally appealing choice of these parameters are
given by Makhlin \cite{makhlin}
\begin{gather}
\alpha = \text{Re } \frac{\text{Tr }^{2} V}{16 \;  \text{det } U}, \\
\beta = \text{Im } \frac{\text{Tr }^{2} V}{16 \;  \text{det } U}, \\
\gamma=  \frac{\text{Tr }^{2} V - (\text{Tr}V)^{2}}{4 \; \text{det } U},
\label{R43a}
\end{gather}
 where $V= W^{\text{T}} W$, $W= Q ^{\dagger} U Q$ and 
  \begin{gather}
 Q= \frac{1}{\sqrt{2}}\left(
 \begin{array}{cccc}
 1   &0   & 0  &  i \\
 0   &i   &1   &  0\\
 0   &i   &- 1  & 0 \\
 1   &0   & 0  & -i 
 \end{array}
 \right).
 \label{R33a}
 \end{gather}
In particular, these parameters uniquely classify equivalence classes
of all 2-qubit generalized permutations and thus the the set CC$(U)$
of all classical counterparts of a given unitary $U$.  For
compactness, it is convinient to first divide the the group of
permutations $S_{4}$ into $6$ cosets with respect to the subgroup of
local permutations $S_{2} \otimes S_{2} = \{ \sigma_{x}^{j} \otimes
\sigma_{x}^{k} | j,k=0,1\}$. These cosets may then be identified with
their respective representatives chosen as follows: $\I$, $\SWAP$,
$\CNOT_{12}$, $\CNOT_{21}$, $\SWAT_{12}$, $\SWAT_{21}$, where $\SWAT
\equiv \SWAP \cdot \CNOT$. The classes of CC$(U)$ are identified in
Table \ref{table1}. There are four non-trivial classes and one empty
class.

% \begingroup
% \squeezetable
\begin{table}[h]

\caption{Complete classification of classical counterparts of 2-qubit
         unitaries
}

\label{table1}
\begin{tabular}{c|c|c|c}
  $\alpha $ & $\beta$ & $\gamma $ & CC(U) \\ \hline\hline 
  &&& \\
  $\alpha $& $ \beta \neq 0$ & $\gamma $ & $ \phi $  \\
  % &&& \\
 $0 $& $  0$ & $1 $ & $\{ \I, \CNOT_{12}, \CNOT_{21} \}$  \\
  % &&& \\
  $0 $& $  0$ & $-1 $ & $\{ \SWAP,\SWAT_{12}, \SWAT_{21}\} $  \\
  % &&& \\
  $\alpha >0$& $0$ & $1 + 2 \alpha  $ & $\{  \I  \} $  \\
  % &&& \\
  $\alpha <0$& $0$ & $-1 + 2 \alpha  $ & $\{\SWAP\} $  
\end{tabular}
\end{table}
% \endgroup

Finally, let us turn to the important question of the source of
quantum speed-up. Although quantum entanglement is believed to be the
key to quantum speed-up, there is no proof that this is indeed the
case. For example, BV problem is a commonly mentioned case where
quantum speed-up seems to be obtained without entanglement \cite{meyer}. The PARITY
problem solution also doesnot use any entanglement.  We believe, that
our notion of genuine quantum speed-up may help clarify the role of
entanglement as a necessary constituent of quantum over classical
algorithmic superiority. In the examples mentioned above we have been
able to show that there is actually no genuine quantum speed-up where
entanglement is absent. Moreover, in examples such as Grover's problem
and the Deutsch-Jozsa problem where entanglement is crucial, we have
not been able to report finding corresponding classical oracles that diminish
the quantum speed-up. Of course, in order to prove the link between
genuine quantum speed-up and entanglement the non-trivial task of
finding all the classical counterparts of an arbitrary multi-qubit
quantum unitary operation must be solved.

Summarizing, we have shown that the common procedure for comparing
quantum and classical oracles is ambiguous. This has led us to
introduce the notion of genuine quantum speed-up which allows reliable
comparison of quantum and classical oracles. As an example, we have
shown that the Bernstein-Vazirani and PARITY problems do not exhibit
genuine quantum speed-up.

\acknowledgments
  
A. W. would like to thank the State Commision for Scientific Research
for financial support under grant no. 0 T00A 003 23.


\begin{references} 

\bibitem{shor}  P. Shor, {\it Proceedings of the  35th Annual Symposium on
  Foundations of Computer Science}, edited by S. Goldwasser (IEEE
  Computer Society Press, Los Alamitos, CA, 1994), p. 124.


\bibitem{parity} E. Farhi, {\it et al.}  Phys. Rev. Lett. {\bf 81}, 5442 (1998). 

\bibitem{deutsch} D. Deutsch, Proc. R. Soc. London, Ser A {\bf 400},
  97 (1985). 




\bibitem{bv} E. Bernstein and U. Vazirani, {\it Proceedings of the 25th ACM
  Symposium on Theory of Computing}, San Diego, CA,(ACM Press, New York
  1993), p. 11; E. Bernstein and U. Vazirani, {\it SIAM J. Comp.} {\bf 26},
  1411 (1997).


\bibitem{exp} P. Londero {\it et al.}, Phys. Rev. A {\bf  69},
  010302 (2004); E. Brainis {\it et al.},  Phys. Rev. Lett. {\bf 90},
  157902 (2003);  J. Siewert and R. Fazio,  Phys. Rev. Lett. {\bf 87},
  257905 (2001); J. Du {\it et al.},  Phys. Rev.  A {\bf 64}, 042306 (2001). 


\bibitem{makhlin} Yu. Makhlin, Quant. Inf. Proc.,  {\bf 1 (4)}, 243 (2003).
\bibitem{cirac} B. Kraus and J. I. Cirac,  Phys. Rev. A {\bf 63},
  062309 (2001). 
\bibitem{zhang} J. Zhang {\it et al.}, Phys. Rev. A  {\bf 67},  042313
  (2003). 

\bibitem{meyer} D. Meyer,  Phys. Rev. Lett. {\bf 85},  2014 (2000). 


\end{references}
\end{document}